\documentclass{moriond}

\usepackage[round]{natbib}   
\def\be{\begin{equation}}
\def\ee{\end{equation}}
\def\bea{\begin{eqnarray}}
\def\eea{\end{eqnarray}}

\newcommand{\Photo}{}

\begin{document}
\vspace*{4cm}
\title{Unraveling the Universe with DESI}

\author{ Mariana Vargas-Maga\~na$^{1}$, 
Brooks,David,D.$^{2}$,
Levi,Michael,M.$^{3}$,
Tarle,Gregory,G.$^{4}$, \\
on behalf of DESI collaboration.}
\address{
\scriptsize $^{1}$Instituto de F\'isica, Universidad Nacional Aut\'onoma de M\'exico, A. P. 20-364, C.d de M\'exico, Mexico.\\
\scriptsize $^{2}$Department of Physics \& Astronomy, University College London, Gower Str, London, WC1E 6BT, UK.\\
\scriptsize $^{3}$Lawrence Berkeley National Laboratory, 1 Cyclotron Road, Berkeley, CA 94720, USA.\\
\scriptsize $^{4}$Physics Department, University of Michigan Ann Arbor, MI 48109, USA.}

\maketitle\abstracts{
The Dark Energy Spectroscopic Instrument (DESI) is a stage IV ground-based dark energy experiment planned to begin operations in 2020. 
In this article, we provided a short review of DESI presented during the conference {\it Recontres de Moriond 2018}. 
DESI will use four different tracers for mapping the universe: from redshift 0.05 up to redshift 1.7 with galaxies and from 2.1 to 3.5 using quasars. DESI will measure a total of 35 million spectra covering regions of universe never explored before, providing a map of large scale structure that will enable major advances in the investigation of cosmic acceleration. The key science goals for DESI are to constrain dark energy and potential deviations of General Relativity using two complementary observables: the  Baryonic Acoustic Oscillations (BAO) and the Redshift Space Distortions (RSD). Additional science goals, such as constraining the sum of neutrino masses and inflation, are expected with the baseline project. DESI  installation started on February 2018 and the current construction of the instrument is on track. The imaging surveys that will serve to determine the targets are currently in the final stages, having achieved  80\% completion, and are expected to be finalized by the end of 2018. The DESI Collaboration is actively preparing for survey operations and science analysis, to be ready for the first light in January 2020.}

\section{Introduction}
One of the most surprising cosmological discoveries in the last decades is the accelerated expansion of the Universe. The first convincing measurement of cosmic acceleration came from observations that type Ia supernovae appeared less luminous than expected in a decelerating Universe \citep{riess1998,perlmutter1998} These observations can be explained by either modifying General Relativity on cosmological scales, or within the framework of the standard cosmological model this implies that 70 \% of the Universe is dominated by a new component called ``dark energy"  the unusual physical property of opposing 
the attractive force of gravity. 

Cosmic acceleration is one of the most important issues to be explored by modern observational cosmology. The implications of the acceleration of the expansion of the Universe have inspired an ambitious experimental program to understand the physics of this phenomenon. Apart from supernovae, there are several other prominent probes that allow us to extract information regarding the equation of state of dark energy. In particular, spectroscopic surveys provided a unique opportunity to explore the expansion history of the Universe as well as to measure the growth of structure through the analysis of the large-scale structure in the Universe. 

During the last 15 years, a succession of photometric and spectroscopic surveys have been operating, all of them driven by the same science goal of decrypting the mysterious cosmic expansion. Just to mention some of the Wide-area spectroscopic survey predecessors of DESI include:  the Sloan Digital Sky Survey-III Baryonic Oscillation Spectroscopic Survey (SDSS-III/BOSS) \citep{Dawson2013}, which operated from 2008-2014 and reported the final results of the experiment in 2016, providing the first 1\% precision measurement of the BAO scale and obtaining the narrowest constraints on the dark energy parameters to date. The successor to BOSS is the extended Baryonic Oscillation Spectroscopic Survey \citep{Dawson2016}, currently the only spectroscopic  survey devoted to cosmology operating as part of the Sloan Digital Sky Survey-IV. Initial results of the analysis with the first two years of data were recently finalized, providing the first clustering measurements with quasars.

Just after eBOSS finishes its program in 2020, the next generation stage IV  ground-based dark energy experiment,  Dark Energy Spectroscopic Instrument (DESI) plans to start operations for 5 years. In this article, we provided a short review of the DESI talk presented during the conference {\it Recontres de Moriond 2018}. The structure of the paper is as follows:  Section ~\ref{sec:concept}  presents the DESI concept. In Section ~\ref{sec:goals}, we present the science goals and requirements of this dark energy experiment, followed by a summary of the
forecast for the key observables of the survey (BAO and RSD) in Section ~\ref{sec:forecast}. The current status of the installation and construction of the instrument is summarized in Section ~\ref{sec:installation} as well as highlights of science preparations currently ongoing.
\section{DESI Concept}\label{sec:concept} 
The DESI instrument is composed of three basic distintive features: 1)  a focal plane assembly with 5000 fiber positioners that will be robotically-controlled;  
2) a new 6-lens wide-field corrector with a field of view (FoV) of 8 square degrees in contrast to the existing corrector that has only 0.5 square degrees FoV; 
3) ten thermally-controlled 3-channel spectrographs over a wavelength from 360 to 980 nm, with a resolution $R=\frac{\lambda}{\Delta \lambda}$ between 2000 and 5500 (depending on the wavelength).
The instrument is planned to be installed in the 4-m Mayall telescope in Kitt Peak, in Arizona. 
The baseline survey is planned to cover 14,000 square degrees  in 5 years, while the requirements for the threshold survey is only 9,000 square degrees. DESI will consist of two programs that depends on the moonlight: the Dark Time and Bright Time. During the dark time, DESI plans to measure spectra from 4 tracers:
\begin{itemize}
\item  4 million Luminous Red Galaxies (LRG) between redshift 0.4 to 1.0. The LRG are massive galaxies that have no star formation and therefore exhibit evolved, red composite spectral energy distributions (SEDs).
DESI will rely on the experience of previous surveys for exploiting the 4000 A break to obtain secure photometric redshifts for LRGs using gri colors. To extend the redshift interval will use the correlation between optical/near-infrared (NIR) color and redshift  generated by the 1.6 $\mu$m bump in the rest frame. 
\item 17.1 million Emission Line Galaxies (ELG) between redshift 0.6 and 1.7. The ELG are galaxies that show high star formation rates, and therefore exhibit strong emission lines from ionized H-II regions around massive stars, as well as spectral energy distributions with a relatively blue continuum, which allows their selection from optical grz photometric bands. The  [OII] doublet in ELG spectra consists of a pair of emission lines separated in rest-frame wavelength by 2.783 A. This wavelength separation of the doublet provides a unique signature, allowing definitive line identification and secure redshift measurements.
\item 1.7 million quasars with redshift from $0.9<z<2.1$  used as direct tracers and 0.7 million Ly-$\alpha$ quasars with redshift between 2.1 and 3.5 for using the Lyman alpha Forest as a tracer of the matter along the line of sight of the quasar. We will use optical photometry combined with WISE infrared photometry in the W1 and W2 bands to select the primary sample of QSOs. The near-infrared allow us to discriminate between quasars and stars, as quasars are brighter at all redshifts compared to stars in the NIR.
\end{itemize} 
During the bright time:
\begin{itemize}

\item 10 million galaxy spectra from 0.05 up to redshift 0.4 for generating a magnitude-limited Bright Galaxy Survey (BGS). The r-band will be used for the selection of the sample of galaxies.
\end{itemize} 
 In total, DESI is expected to measure 35 million spectra of galaxies and quasars. All DESI target samples will be selected using optical grz-band photometry from ground-based telescopes \citep{arjun2018} and near-infrared photometry from the WISE satellite. The DESI survey relies on pre-imaging for target selection. In the following section, I briefly describe the imaging surveys that will be used for DESI.
  \begin{figure}
\centerline{\includegraphics[width=0.4\linewidth]{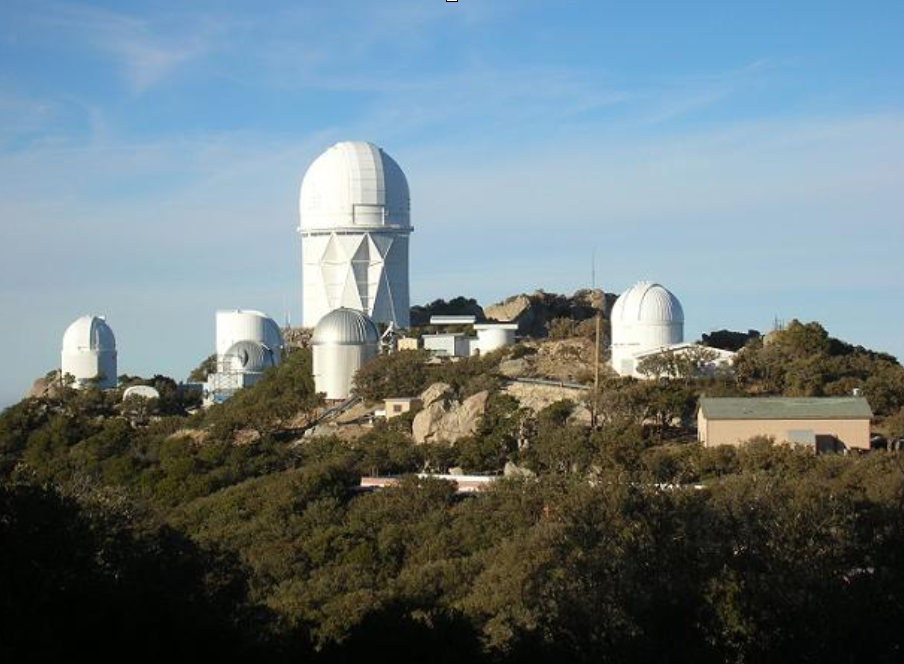}\includegraphics[width=0.4\linewidth]{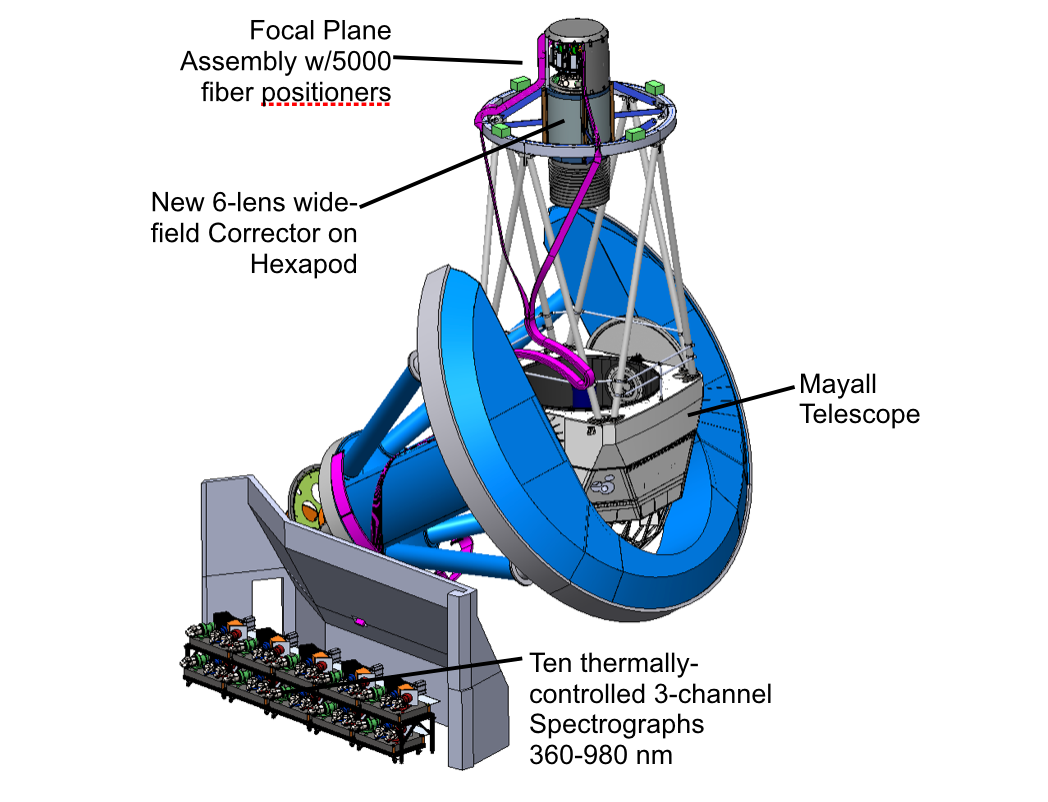}} 
\caption[]{ Left panel: 4-m Mayall telescope in Kitt Peak, in Arizona. Right panel:DESI instrument.}
\label{fig:error}
\end{figure}
 \subsection{Imaging Surveys}\label{subsec:imaging}
The DESI Collaboration is participating closely in 3 major imaging surveys. 
\begin{itemize}
\item Dark Energy Camera Legacy Survey (DECaLS). This survey uses Blanco DECam 4-m telescope situated  Cerro Tololo Inter-American Observatory. This survey will provide grz imaging over 9000 square degrees region in the DESI footprint at DEC$<=$+34 degrees of the North Galactic Cap (NGC) and South Galactic Cap (SGC). DECaLS survey will observe the sky in three passes. This survey started in 2014 and  is 71\% complete in the NGC and 53\% complete in the SGC as reported in May 2018. It is expected to be completed in January 2019.
\item Beijing-Arizona Sky Survey (BASS). This survey uses the Bok 2.3m telescope situated in Kitt Peak (just next to the Mayall telescope). This telescope will provide gr imaging over a 5000 square degree region of the NGC lying at DEC$<=$+34 degrees. This survey started in 2015 and it is 66\% completed in the g band and 65\% in the r band as reported in May 2018. The survey is expected to be completed in August 2018.
\item Mayall z-band Legacy Survey (MzLS). This survey uses the Mayall MOSAIC-3 Telescope in Kitt Peak. This survey  provides z  imaging over a 5000 square degree region region of the North Galactic Cap (NGC) lying at dec $>= +$34 degrees. The BASS survey tiles the sky in three passes. This survey started in 2016 and it was completed in February 2018. 
\item Wide-field Infrared Survey Explorer (WISE). WISE is an infrared imaging obtained from a satellite. WISE conducted an all-sky survey in four bands centered at 3.4, 4.6, 12 and 22 $\mu m$ (known as W1, W2, W3 and W4). DESI target selection utilizes the two wavelength bands at W1 and W2. This survey was  completed in 2017.
\end{itemize}
The DESI analyses will be performed separately in each of the three regions of the DESI footprint: the NGC at DEC $> +$34 degrees, the NGC at DEC$<$+34 degrees, and the South Galactic Cap (SGC). 
Completion of observing for imaging surveys December 2018.
\section{Sciences Goals and Requirements}\label{sec:goals}
As described in the Science Requirements\footnote{\it ``The primary scientific purpose of the Dark Energy Spectroscopic Instrument (DESI) is to probe the origin of cosmic acceleration by employing the baryon acoustic oscillation (BAO) technique as a ``standard ruler" to measure the expansion history of the Universe with improved precision, extending the measurement back to redshifts of z =3.7. "(Science Requirements)}, 
DESI will explore some of the most fundamental questions of modern cosmology, the acceleration of the cosmic expansion. This phenomenon could driven by, either a modification to General Relativity or a new form of energy, known as dark energy.
Since there are a large number 
of alternative models, the phenomenological way of proceeding is to describe the cosmic acceleration as generated by a new energy component in the energy-momentum tensor that in such a way that can be modeled as a perfect fluid , parameterize its effective state equation, 
and determine the limits of the parameters of the equation of state to be able to discriminate between the different alternative models. The most used parameterization by the observational community, is the CPL with a State Equation (EdE):
\begin{equation}
w(z)=w_0+w_a (1-a),
\end{equation}
where $a$ is the scale factor, $w_0$ is the value of the dark energy equation of state at present, and $w_a$ is the derivative describing 
its time evolution.

The key goal of DESI is to achieve constraints on the dark energy equation of state parameters with a precision of a Stage IV Dark Energy Experiment as defined in the Dark Energy Task Force report \citep{DETF}.
The precision is quantified using the Dark Energy Task Force  (DEFT) Figure of Merit (FoM).\
 The FoM\footnote{We can define the pivot $w_p$ such as the covariance matrix is diagonal and then $FOM=1/[\sigma_{w_a}*\sigma_{w_p}] $.}  is defined as:
 \begin{equation}
FoM  =\frac{1}{\sqrt{det[COV(w_0,w_a)]} } 
\end{equation}
The figure of merit accounts for the improvement expected for different stages of the dark energy experiments. The DETF defined 4 stages: 
 Stage II experiments were those in progress at the time of the report as SDSS-I-II. Stage III experiments are those improving upon the Stage II FoM by at least a factor of 3, and include the BOSS and eBOSS experiments.
 The figure of merit of the parameters of the equation of state of the dark energy  expected for DESI should be 3 times better than the stage III experiments, i.e 10 times better than stage II experiments.
Table \ref{tab:detf} shows an extract of Table 2.9 from the DESI Experiment Part I: Science,Targeting, and Survey Design \citep{desi} where the Dark Energy Task Force (DETF) Figures of Merit for different scenarios are quoted. The minimal scenario considering only galaxies in the baseline survey accomplishes the criteria of stage IV dark-energy experiments.

\begin{table}[t]
\caption[]{Dark Energy Task Force (DETF) Figures of Merit for DESI}
\label{tab:detf}
\vspace{0.4cm}
\begin{center}
\begin{tabular}{|l|c|}
\hline
Survey & FoM  \\

 \hline
BOSS BAO & 37 \\
DESI 14k galaxy BAO & 133 \\
DESI 14k galaxy and Ly-$\alpha$ forest BAO& 169 \\
DESI 14k galaxy BAO +gal broadband to $k<0.1 h $Mpc$^{-1}$&332  \\
DESI  9k galaxy BAO & 95\\
DESI  9k galaxy and Ly-$\alpha$ forest BAO &121\\
DESI 9k galaxy BAO +gal broadband to $k<0.1 h $Mpc$^{-1}$&229 \\
\hline
\end{tabular}
\end{center}
\end{table}

For achieving the precision required for a stage IV dark energy experiment, DESI is designed to use two robust techniques: the Baryon Acoustic Oscillations (BAO) and Redshift Space Distortions (RSD), both in one survey. 
\begin{itemize}
\item The BAO feature is an enhancement at $\sim150$ Mpc in the two-point correlation function of matter in the Universe. This corresponds to the maximum distance travelled by acoustic waves in the matter-radiation fluid during the period from matter/radiation equality to their decoupling at $z \sim1000 $  and then stretched by expansion of the Universe. An excess density, which appears today at a radius of 150 Mpc, was left after decoupling around each dark-matter density peak. We therefore expect to find today an excess probability to find matter separated by this distance (in comoving coordinates). This feature can be seen as a standard ruler allowing us to study the history of the expansion of the Universe and infer cosmological information. The physics underlying the acoustic oscillations is well understood and the value of this standard ruler has been measured with exquisite accuracy at a redshift of $z \sim1100 $ using CMB data. The apparent BAO distance scale measured from observations at different redshifts leads to measurements of the Hubble parameter H(z) and the angular diameter distance $D_A(z) $, which are related to the cosmological parameters, especially to the dark energy parameters.
\item For the study of cosmic acceleration, an observable complementary to distance measurements are measurements of the growth of the structure\footnote{To illustrate the complementarity of the growth of structure and the BAO, we notice that the FoM of the analysis including broadband of the galaxy clustering is ~3 times better than only using BAO information, thus the RSD analysis  reduces the uncertainty on the dark energy equation of state parameters.}. In general, the greater the acceleration of the expansion, the greater the suppression of the growth of the structure. The growth function tells us how the amplitude of the cluster of galaxies scales with cosmic time. Measurements of the growth function provide information about dark energy  and even in the scenario without dark energy, these measurements allow us to test whether General Relativity describes the laws of physics on a large scale. Due to the gravitational growth, the galaxies tend to be attracted towards the over-dense regions; consequently the observed redshifts are distorted by these peculiar velocities in the direction of the line of vision. These distortions generate an increase in clustering along the line of sight compared to the perpendicular direction. The measurement of relative clustering along and perpendicular to the line of vision leads to measurements of the logarithmic rate of growth of the structure: $f (a) \sigma_8 (a)$. 
\end{itemize}
The survey is designed to measure the distance scale from BAO with 0.28\% precision from $0 < z < 1.1$  and 0.39\% precision from $1.1 < z < 1.9$; additionally, the precision expected for the measurements of the Hubble Parameter is 1.05\% at $1.9 < z < 3.7$ from anisotropic BAO analysis. 
To achieve this accuracy, the systematic errors from instrument and observational effects must not exceed 0.16\% for $D_A(z)$ and 0.26\% for $H(z)$.
The gravitational growth measurements' precision is expected to be  $< 1\% $ at $0.5 < z < 1.4 $ using RSD. 

Additionally, there are secondary science goals that could be achieved with the baseline plan, three in particular: 
\begin{enumerate}
\item Inflation constrains can be obtained from the k-dependence of the broadband power; the spectral index $n_s$ of primordial perturbations is expected to be measured with $\sigma_{n_s}=0.0025 $ and its running with wavenumber $\alpha_s$ is expected to be measured with $\sigma_{\alpha_s}= 0.004$\footnote{Using the broadband galaxy power spectrum  only of galaxies out to $k_{max} = 0.1$}.
\item Primordial non-Gaussianity can also be explored through the large scale clustering; usually, the type of non-gaussianity studied is the non-local type parametrized by $f_{NL}$, the projection on the precision on the $f_{NL}$ is $\sigma_{f_{NL}}=5$. 
\item Neutrino constrains, in particular, the measurement of the sum of neutrino masses to $< 0.02 $eV with an uncertainty of 0.020 eV (for $k_{max}< 0.2 h Mpc ^{-1}$), and the study of the neutrino hierarchy\footnote{With DESI measurement of the absolute mass scale, we can determine which hierarchy is preferred by the data, given that the normal and inverted hierarchy imply different value for the sum of masses.} \citep{font2014}.
\end{enumerate}
In next section, I will provide the error projections for the baseline survey in several redshift slices for the different tracers only for the primary goals: the BAO and RSD observables.
\section{Forecast for BAO and RSD observables}\label{sec:forecast}
DESI will provide at least an order of magnitude improvement over BOSS both in the comoving volume and the number of galaxies measured. 
The results of performing a Fisher matrix formalism for estimating the parameter-constraining power of the finished survey are shown in Table 2.3 and Table 2.4  from the DESI Experiment Part I: Science,Targeting, and Survey Design \citep{desi} for the baseline survey and for the threshold survey respectively. 
These tables quote errors on the transverse and radial BAO scales as errors on $D_A(z)/s$ and $H(z)s$, respectively, where $s$ is the BAO length scale, 
and also quote errors on an isotropic dilation factor $R/s$, defined as the error one would measure on a single parameter that rescales radial and transverse directions by equal amounts. In this review, I  show in the left panel of Figure \ref{fig:error} the error projections only for the Hubble parameter on the different redshift bins with the different tracers using the numbers from the tables mentioned previously. Also, in the right panel of Figure \ref{fig:error}, I show the comparison of the DESI error projections with other experiments present and future to highligth the constraining power of DESI. 
\begin{figure}
\centerline{\includegraphics[width=0.3\linewidth]{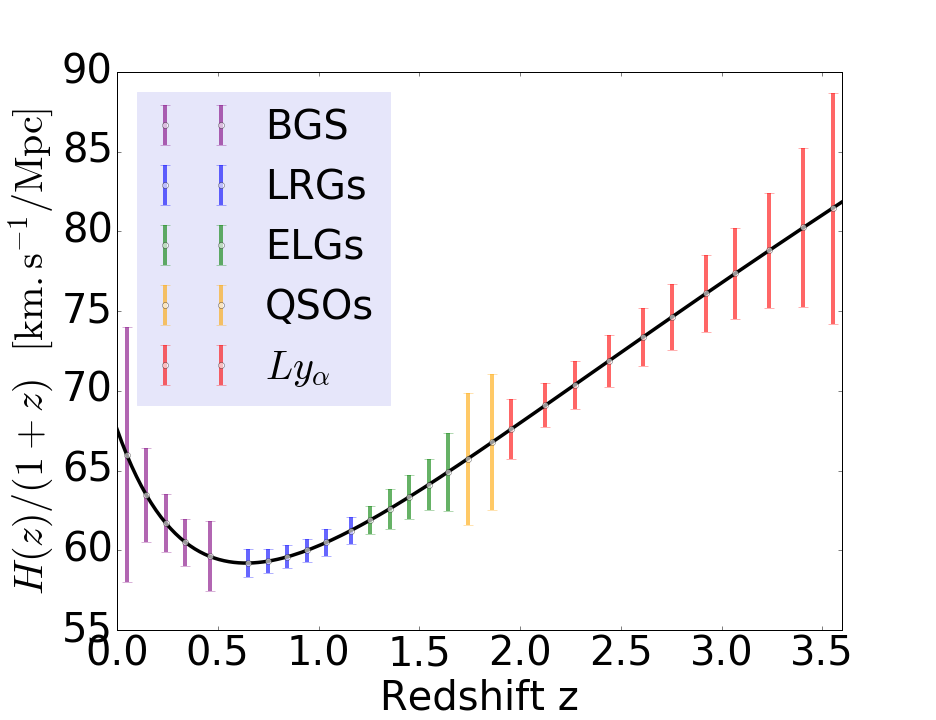}\includegraphics[width=0.3\linewidth]{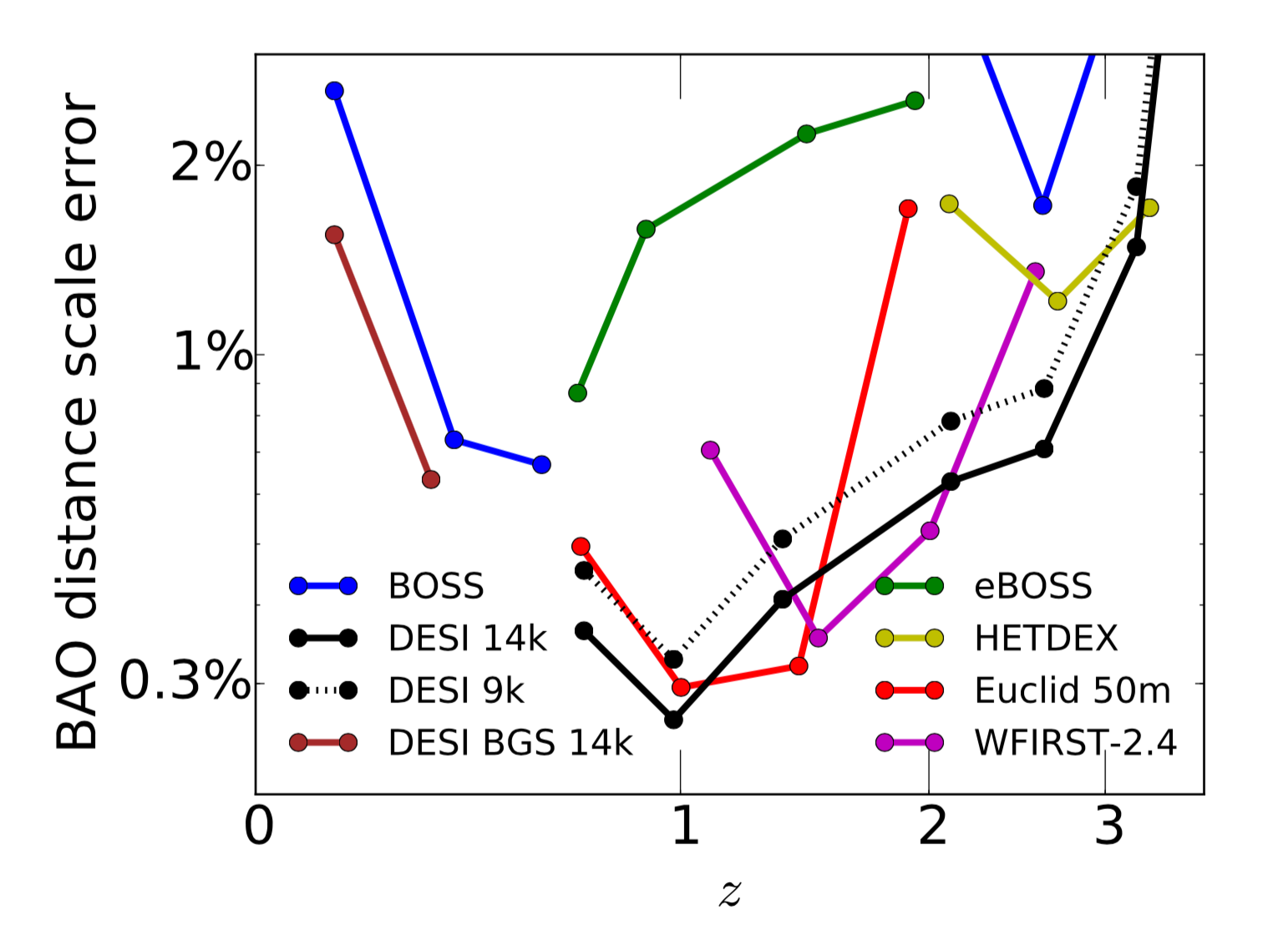}}
\caption[]{Left panel: Error projections for the Hubble parameter for DESI in several redshift bins corresponding to different tracers corresponding to the Tables 2.3, 2.4 of the DESI Experiment Part I: Science,Targeting, and Survey Design \citep{desi}. Right panel: Comparison of the DESI error projections with other experiments present and future [Right panel adopted from the DESI Experiment Part I: Science,Targeting, and Survey Design \citep{desi}].}
\label{fig:error}
\end{figure}

\begin{figure}
\centerline{\includegraphics[width=0.45\linewidth]{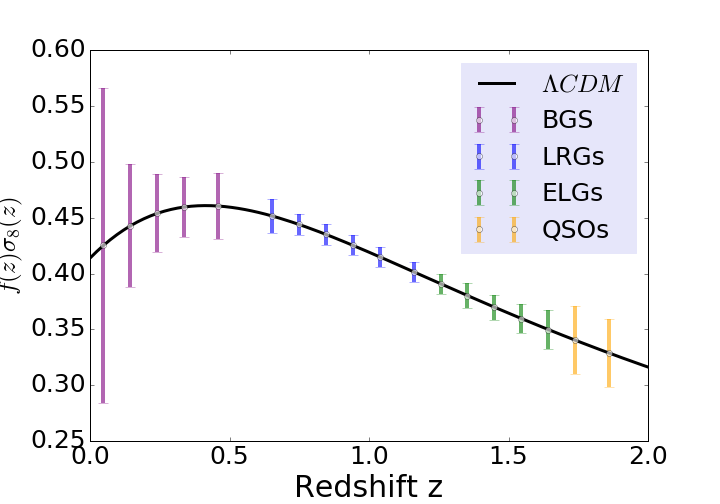}\includegraphics[width=0.4\linewidth]{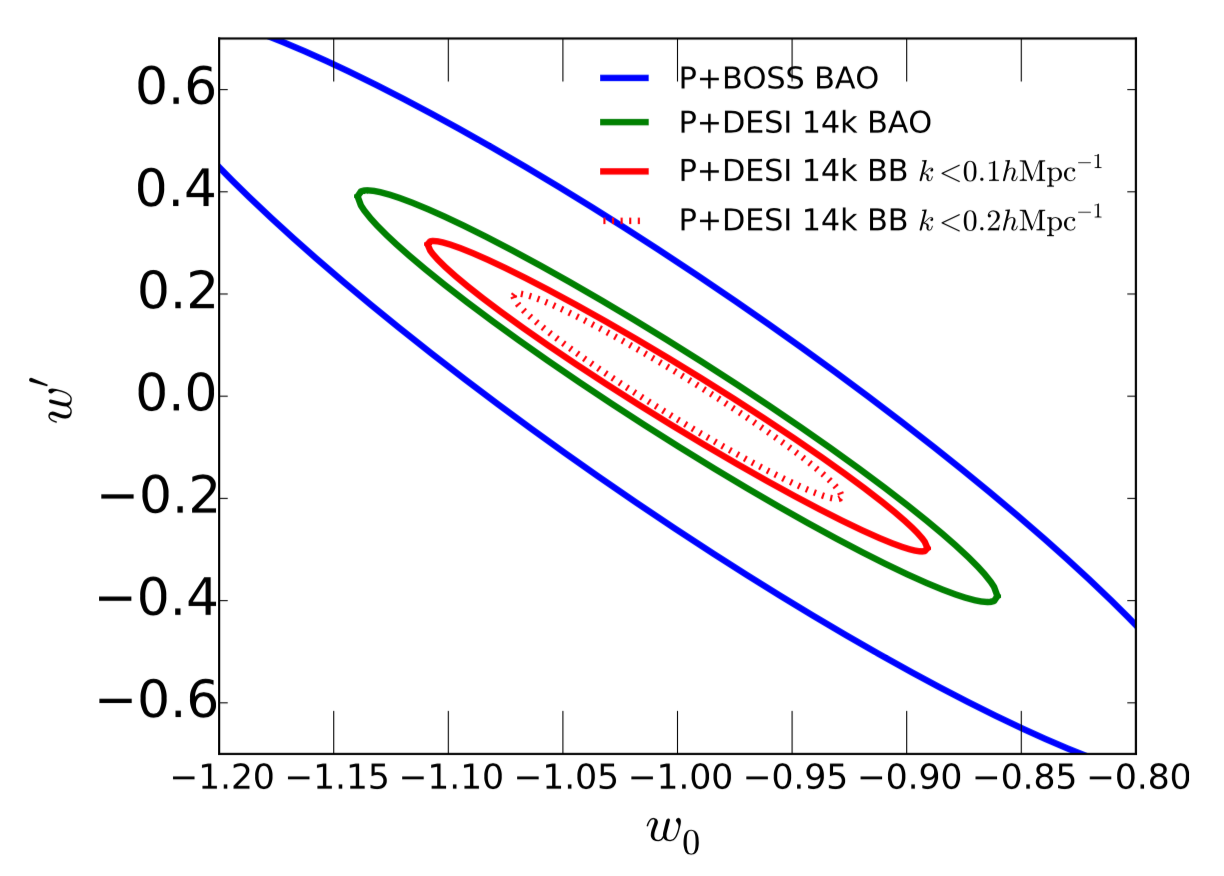}}
\caption[]{Left panel: Error projections for the $f(z) \sigma_8(z)$ for DESI in several redshift bins corresponding to the different tracers corresponding to the Tables 2.3, 2.4 of the DESI Experiment Part I: Science,Targeting, and Survey Design \citep{desi}. Right panel: Comparison of the DESI 1-$\sigma$ contours in the $w_0-w'$ plane with other experiments present and future.  [Right panel figure adopted from the DESI Experiment Part I: Science,Targeting, and Survey Design \citep{desi}]} 
\label{fig:growth}
\end{figure}

The redshift-space distortions can effectively constrain two parameter combinations, $b(z) \sigma_8 (z)$ and $f(z) \sigma_8(z)$. For the forecast numbers, the large-scale broadband power was used up to some quoted $k_{max}$. We quote in this review only the numbers for the baseline 14K survey and considering only the $k_{max}=0.1 h$ Mpc$^{-1}$, which corresponds roughly to the performance of current analyses. 
 The left panel of Figure~\ref{fig:growth} shows the rate of growth of the structure, $f$, as a function of the redshift, for an LCDM model together with the galaxy and quasar error projections (including the Bright Galaxy Survey). DESI will significantly extend in redshift the growth factor measurements with better precision than current stage III experiments.
 To finalize this section, I present in the right panel of Figure~\ref{fig:growth} the 1-$\sigma$ contours in the $w_0-w'$ plane  expected for DESI and the comparison with past and present Stage III dark energy experiments.
 
\section{Construction Progress and Science Preparation}\label{sec:installation}

\subsection{Construction Progress and Timeline}
The construction progress and installation is a 60\% complete which puts it roughly on schedule. The Mayall telescope is currently being prepared for the installation of the new optical corrector. The installation started on February 12, 2018 and is scheduled to be completed in the summer of 2018.
The next major step in the installation is the DESI focal plane system,  that is scheduled to begin installation in March 2019.
To fill the gap, between the two major steps of installation,  a commissioning instrument was designed and fabricated so it can be installed on the Mayall Telescope in the fall of 2018,  after the new optical corrector is  installed. This commissioning instrument will serve to test the imaging capability and will allow to measure the image quality over the full DESI field of view. 

Commissioning of the complete instrument is scheduled  to start in May 2019 and planned to be completed by September 2019. The  goal of commissioning is to verify that the DESI system can take survey-quality data. The next step in the timeline is the survey validation, which is planned  for September 2019. The goal of validation is to connect DESI requirements to the quality assurance and verification tests that will ensure that the data systems meet those requirements.
Finally, the science survey is planned to start in January 2020. The survey will be organized into data assemblies for internal data release, scoped to be a stable and accessible foundation for our collaboration science analyses, including raw and reduced data, large-scale structure catalogs, mock catalogs, and suitable documentation. The first data assembly that will include data up to June 2020 and is planned to occur in January 2021. Further data assemblies and cosmology analyses will occur on 1-year cycles.The first data assembly for cosmology analysis is planned for April 2022. 
The end of the survey is planned to be January 2025.

\subsection{Science Preparacion}\label{subsec:scienceprep}

The DESI Collaboration is actively preparing for survey operations and science analysis. We have a large collaboration (more than 500 members) with rich experience in cosmology and survey astronomy. The
working groups have developed detailed plans to achieve their science objectives and are now working to carry those out. The main document of this is the  Science Requirements Document. 
The collaboration has been actively preparing 
DESI operations in 2019, as demonstrated  through the creation elaboration of the following documents :
\begin{itemize}
\item Science Readiness Plan, which itemizes the key tasks and schedule needed to prepare for running and validating the survey and doing the first set of science analysis.
\item Baseline Survey Strategy, which describes how the dark-time survey should proceed.
\item Bright Galaxy Survey Plan, which describes the bright time survey and how it should proceed.
\item Target Selection Baseline, which describes the current state of the target selection algorithms.
\item Baseline Survey Validation Plan, which lays out a plan for the SV period, the main goal of which is to validate the survey strategy and confirm that we are ready to begin.
\item Cosmological Simulations Requirements and Plan, which describes what simulations are required for the core science as well as what simulations the collaboration should produce to reach its broader science goals.
\item Imaging Validation Plan, which describes how we will determine that the imaging meets requirements.
\end{itemize}
I would like to finalize this review with a quote from David Schlegel, that I think reflects accurately the mood of the collaboration at this moment, that we are a two years from the first light of DESI.
{\it ``For the past 13 years, we've had a simple model of how dark energy works. But the truth is, we only have a little bit of data, and we're just beginning to explore the times when dark energy turned on. If there are surprises lurking out there, we expect to find them."}

\section*{Acknowledgments}
\tiny{ MV is partially supported by Programa de Apoyo a Proyectos de Investigaci\'on e Innovaci\'on Tecnol\'ogica (PAPITT) No  IA102516 and Proyecto Conacyt Fronteras No 281. 

This research is supported by the Director, Office of Science, Office of High Energy Physics of the U.S. Department of Energy under Contract No. 
DE-AC02-05CH1123, and by the National Energy Research Scientific Computing Center, a DOE Office of Science User Facility under the same 
contract; additional support for DESI is provided by the U.S. National Science Foundation, Division of Astronomical Sciences under Contract No. 
AST-0950945 to the National Optical Astronomy Observatory; the Science and Technologies Facilities Council of the United Kingdom; the Gordon 
and Betty Moore Foundation; the Heising-Simons Foundation; the National Council of Science and Technology of Mexico, and by the DESI 
Member Institutions: Aix-Marseille University;  Argonne National Laboratory; Barcelona Regional Participation Group; Brookhaven National Laboratory; 
Boston University; Carnegie Mellon University; CEA-IRFU, Saclay; China Participation Group; Cornell University; Durham University;  \'Ecole Polytechnique 
F\'ed\'erale de Lausanne; Eidgenossische Technische Hochschule, Zurich;  Fermi National Accelerator Laboratory;  Granada-Madrid-Tenerife Regional 
Participation Group; Harvard University; Korea Astronomy and Space Science Institute; Korea Institute for Advanced Study; Institute of Cosmological  
Sciences, University of Barcelona; Lawrence Berkeley National Laboratory; Laboratoire de Physique Nucléaire et de Hautes Energies; Mexico Regional 
Participation Group; National Optical Astronomy Observatory; Ohio University; Siena College; SLAC National Accelerator Laboratory;  Southern Methodist University; 
Swinburne University; The Ohio State University; Universidad de los Andes; University of Arizona; University of California, Berkeley; University of California, 
Irvine; University of California, Santa Cruz; University College London; University of Michigan at Ann Arbor; University of Pennsylvania; University of Pittsburgh; 
University of Portsmouth; University of Rochester; University of Queensland; University of Toronto; University of Utah; University of Zurich; UK Regional Participation Group; Yale University. The authors are honored to be permitted to conduct astronomical research on Iolkam Du'ag (Kitt Peak), a mountain with particular significance to the Tohono O'odham Nation.  For more information, visit desi.lbl.gov.}

%


\bibliography{moriond1}
%
%
%


\end{document}